Reply to Comments on "Stability of Tsallis entropy and instabilities of Rényi and

normalized Tsallis entropies: A basis for $q$-exponential distributions"

Sumiyoshi Abe

*Institute of Physics, University of Tsukuba, Ibaraki 305-8571, Japan*

**Abstract**

Bashkirov's comments on the paper [S. Abe, Phys. Rev. E **66**, 046134 (2002)] are all refuted. In addition, it is discussed that the Rényi entropy is irrelevant to generalization of Boltzmann-Gibbs statistical mechanics for complex systems.

In a recent short note [1], Bashkirov has criticized the work in Ref. [2] by considering the following three points: the concept of the thermodynamic limit, comparison of the values of the Tsallis entropy and the Rényi entropy, and a possible role of the Rényi entropy in generalizing Boltzmann-Gibbs statistical mechanics. Since Bashkirov's discussion seems to represent typical misunderstandings of the issues, it may be meaningful to reply to his comments in order to make further clarification.

First of all, Bashkirov mixes the problem of the order to be taken for the thermodynamic limit in ordinary statistical mechanical calculations with the order of limits in the Lesche-stability condition [3]. Here, a statistical entropy functional $C[p]$ of a probability distribution $\{p_i\}_{i=1,2,\cdots,W}$ is called Lesche-stable if the following condition is satisfied:

$$(\forall\, \varepsilon > 0)\, (\exists\, \delta > 0) \left( \|p - p'\|_1 < \delta \Rightarrow \left| \frac{C[p] - C[p']}{C_{\max}} \right| < \varepsilon \right), \qquad (1)$$

for any value of $W$, where $\{p'_i\}_{i=1,2,\cdots,W}$ is an arbitrary deformation of $\{p_i\}_{i=1,2,\cdots,W}$, $\|A\|_1$ the $l^1$ norm of $A$, and $C_{\max}$ the maximum value of $C$. Specifically, one is concerned with the case of the thermodynamic limit, $W \to \infty$. As rigorously proved in Refs. [2,3], the Rényi entropy [4], $S_q^{(R)}[p] = (1-q)^{-1} \ln \sum_{i=1}^{W} (p_i)^q$, and the normalized Tsallis entropy [5,6], $S_q^{(NT)}[p] = (1-q)^{-1} \left[ 1 - 1 / \sum_{i=1}^{W} (p_i)^q \right]$, are not Lesche-stable,

whereas the Tsallis entropy [7], $S_q^{(T)}[p] = (1-q)^{-1} \left[ \sum_{i=1}^{W} (p_i)^q - 1 \right]$, is Lesche-stable. Here, $q$ in these expressions is the positive entropic index, and all of the above three quantities converge to the Boltzmann-Gibbs-Shannon entropy, $S[p] = -\sum_{i=1}^{W} p_i \ln p_i$, in the limit $q \to 1$. Also, the unit is used, in which Boltzmann's constant becomes unity.

Bashkirov mentions that the thermodynamic limit, $W \to \infty$, has to be taken at the end of calculations. This is true if the calculations are about macroscopic thermodynamic quantities at strict equilibrium. However, the Lesche-stability condition has nothing to do with calculations of macroscopic thermodynamic quantities at equilibrium. Instead, it is concerned with the analytic property of an entropic functional under consideration. More precisely, it defines uniform continuity of such a functional. For physical entropy relevant to ordinary statistical mechanics, two limits, $W \to \infty$ and $\delta \to +0$, may commute. However, the order, $\delta \to +0$ after $W \to \infty$, is nontrivial, in general. Such an order is actually of central interest for studies of statistical mechanics of complex systems in nonequilibrium stationary states [8,9]. There, $\delta \to +0$ corresponds to the long-time limit, $t \to \infty$, describing relaxation of $\{p'_i\}_{w=1,2,\cdots,W}$ to $\{p_i\}_{i=1,2,\cdots,W}$ representing a nonequilibrium stationary state. The order, $t \to \infty$ after $W \to \infty$, is at the heart of nonextensive statistical mechanics [8,9], whereas the order, $W \to \infty$ after $t \to \infty$, is nothing but the ordinary Boltzmann-Gibbs case. Thus, the definition in Eq. (1) correctly reflects nontriviality of the order of the limits. It is also connected to experimental robustness [10] of the quantity, *C*. Usually, what is experimentally observed is not *C* itself but the distribution of the values of a physical quantity under consideration. Repeating the same experiment to measure the values of the same physical quantity, an experimentalist will obtain a distribution, which may be slightly different from that observed previously. *C* should not change drastically for two slightly different distributions, $\{p_i\}_{i=1, 2, \cdots, W}$ and $\{p'_i\}_{w=1, 2, \cdots, W}$, irrespectively of the value of *W*. In a recent paper [11], Lesche has further developed a discussion about the fact that the Rényi entropy cannot be related to observables.

Secondly, Bashkirov stresses the relation in Eq. (3) in Ref. [1]. It is mathematically true but physically irrelevant, since it is just comparing the bare values of two different quantities of two different theories. Only comparison of values of each individual entropy is meaningful. In addition, the quantities examined by Bashkirov are not bounded functionals, in general and accordingly the divisions by their maximum values, as in Eq. (1), are essential.

Thirdly, it is unlikely that the use of the Rényi entropy for generalizing Boltzmann-Gibbs statistical mechanics makes sense. There are at least two important issues, here. One is that the microcanonical structure of the Rényi-entropy-based theory is identical to that of Boltzmann-Gibbs theory since both the Boltzmann-Gibbs-Shannon entropy and the Rényi entropy yield the same value, $\ln W$, for the equiprobability, which is in contrast, for example, to the corresponding value of the Tsallis entropy, $(W^{1-q} - 1)/(1 - q)$. Therefore, no differences appear at the level of macroscopic thermodynamics. The other is concerned with concavity. The concavity condition

$$S_q^{(R)}[\lambda p^{(1)} + (1-\lambda) p^{(2)}] \geq \lambda S_q^{(R)}[p^{(1)}] + (1-\lambda) S_q^{(R)}[p^{(2)}] \quad (0 \leq \lambda \leq 1), \quad (2)$$

holds if $0 < q < 1$, but $S_q^{(R)}[p]$ possesses neither concavity nor convexity if $q > 1$. [12]. Therefore, $S_q^{(R)}[p]$ with $q > 1$ fails to measure the degree of lack of information [13], showing that it cannot be an entropy. Accordingly, one has to limit oneself to $S_q^{(R)}[p]$ with $0 < q < 1$. This, however, turns out to lead necessary to an "opportunistic" treatment, implying a change in the definition of the expectation value. This can be seen as follows.

If the ordinary expectation value

$$<Q> = \sum_{i=1}^{W} Q_i \, p_i \quad (3)$$

is employed for a physical quantity $\{Q_i\}_{i=1, 2, \cdots, W}$, then the maximum Rényi-entropy method, $\delta \left\{ S_q^{(R)}[p] - \alpha \left( \sum_{i=1}^{W} p_i - 1 \right) - \beta \left( \sum_{i=1}^{W} Q_i p_i - <Q> \right) \right\} = 0$ (with the Lagrange multipliers, $\alpha$ and $\beta$, associated with the normalization condition and the expectation value, respectively), yields up to the normalization constant the following stationary distribution:

$$\tilde{p}_i \sim \left[1 - \frac{q-1}{q}\beta\left(Q_i - <Q>^\sim\right)\right]_+^{1/(q-1)}, \qquad (4)$$

where $[a]_+ \equiv \max\{0, a\}$ and $<Q>^\sim$ is the value of $<Q>$ in Eq. (3) calculated in terms of the stationary distribution in Eq. (4) in a self-referential manner. On the other hand, if the *q*-expectation value [14-16]

$$<Q>_q = \frac{\sum_{i=1}^{W} Q_i (p_i)^q}{\sum_{j=1}^{W} (p_j)^q} \qquad (5)$$

is used, then the corresponding stationary distribution reads

$$p_i^* \sim \left[1 - (1-q)\beta\left(Q_i - <Q>_q^*\right)\right]_+^{1/(1-q)}, \qquad (6)$$

where $<Q>_q^*$ stands for the value of $<Q>_q$ calculated in terms of the stationary distribution in Eq. (6) itself. In Eqs. (4) and (6), the same symbol, $\beta$, is used for the Lagrange multipliers, but it will not cause any confusion. Now, recall that $S_q^{(R)}[p]$ is an entropy if and only if $0 < q < 1$. Therefore, $\tilde{p}_i$ in Eq. (4) describes an asymptotically power-law distribution of the Zipf-Mandelbrot type, whereas $p_i^*$ in Eq. (6) is support-compact with the cut-off at $Q_i^{\max} = [1 + (1-q)\beta <Q>_q^*] / [(1-q)\beta]$. Very importantly, the distributions of both of these two types are observed in nature (see a lot of examples collected in the list given at the URL; http://tsallis.cat.cbpf.br/TEMUCO.pdf). Thus, one concludes that, to describe the distributions of the both types, the definition of the expectation value has to be changed depending upon circumstances, losing the possibility of constructing a unified framework. In this respect, we may emphasize that up to now only the Tsallis entropy can lead to a coherent description of the distributions of the both types mentioned above.

Finally, we also point out that the Rényi entropy is not the one that may be relevant to nonlinear dynamical systems prepared at the edge of chaos. The Rényi entropy never yields the constant entropy production rate at the edge of chaos, in contrast to the fact [17-19] that the Tsallis, gamma [19], $\kappa$- [20], and quantum-group [21] entropies do.

In conclusion, I have replied to all of Bashkirov's comments on Ref. [2], and have also discussed why the Rényi entropy is not suitable for generalizing Boltzmann-Gibbs statistical mechanics.

I would like to thank A. K. Rajagopal, A. Robledo, and C. Tsallis for discussions, which were not limited to the present subject. This work was supported in part by the Grant-in-Aid for Scientific Research of Japan Society for the Promotion of Science.